\documentclass{jaa}

\usepackage{array}
\usepackage{graphicx}
\usepackage{amsmath,mathtools}
\usepackage{wasysym}
\usepackage{xcolor}
\usepackage[cmr]{emf}
\begin{document}\sloppy

%\nocite{*} % added by Shashikant
%%paper title
%%For line breaks \\ can be used within title
%\title{Slow Accretion of Helium Rich Matter onto C-O White Dwarf: \\ Does Composition of WD Matter?}
%\title{Effect of WD Composition on Helium Ignition During Slow Accretion}
\title{Accreting White Dwarfs: Effect of WD Composition on Helium Ignition During Slow Accretion}

%%author names are separated by comma (,)
%%use \and before the last author name
%%use a * along with the number separated by comma
%% for the  author for correspondence
%%\textsuperscript{number} is used for affiliation
%%\affilOne, \affilTwo etc., upto \affilTwentyfive is possible
%%Please note the first letter after \affil is capitalised in the command
%%

\author{Harish Kumar\textsuperscript{1}, Abhinav Gupta\textsuperscript{2}
, Siddharth Savyasachi Malu \textsuperscript{3}
         , and Shashikant Gupta\textsuperscript{1,*}}
\affilOne{\textsuperscript{1}G D Goenka University Gurugram, India.\\
\textsuperscript{2}Amity University Gurgaon, India.\\
\textsuperscript{3}Indian Institute of Technology Indore, India.\\}
%https://orcid.org/0000-0002-3702-7069(Harish Kumar)
%https://orcid.org/0000-0003-2923-9245(Dr. Shashikant)
%https://orcid.org/0000-0002-1842-961X(Siddharth)

%%escape two column mode for title, affiliation and abstract
%%by giving \twocolumn command as shown

\twocolumn[{

\maketitle

%%include \corres to print the corresponding author Email id
\corres{shashikantgupta.astro@gmail.com}

%%include \msinfo for
%%manuscript information such as
%%received, revised and accepted dates
%%
\msinfo{}{}

%%abstract
\begin{abstract}\\
Understanding the explosion mechanism of type Ia supernova is among the most challenging issues in astrophysics. Accretion of matter on a carbon-oxygen white dwarf from a companion star is one of the most important keys in this regard. Our aim is to study the effects of WD composition on various parameters during the accretion of helium rich matter at a slow rate. We have used the computer simulation code “Modules for Experiments in Stellar Astrophysics” (MESA) to understand the variations in the properties such as specific heat (\textit{$C_P$}) and degeneracy parameter (\textit{$\eta$}). The profile of specific heat shows a discontinuity and that of the degeneracy parameter shows a dip near the ignition region. As expected, the size of WD decreases and \textit{g} increases during the accretion. However, a red-giant-like expansion is observed after the rapid ignition towards the end. Our study explains the reason behind the delay in onset of helium ignition due to the difference in carbon abundance in a CO-WD. We find that white dwarfs of the lower abundance of carbon accrete slightly longer before the onset of helium ignition.
%...............
%\textbf{Background}: Understanding the explosion mechanism of type Ia supernova is one of the most challenging issues in astrophysics. Accretion of matter on a carbon-oxygen white dwarf from a companion star is one of the most important keys in this regard.
%\\
%\textbf{Objectives}: We wish to study the effects of composition on various properties of WD during the accretion of helium at a slow rate. 
\\
%\textbf{Methods}: We have used the computer code “Modules for Experiments in Stellar Astrophysics” (MESA) for the study of accretion. Variations in the properties of WD such as specific heat and degeneracy profile have been explored in our study.  
%\\
%\textbf{Findings}: The profile of specific heat \textit{$C_P$} and the degeneracy parameter \textit{$\eta$} have been presented. A discontinuity in the specific heat and a dip in the degeneracy parameter can be seen near the ignition region.
%As expected, the size of WD decreases and g increases during the accretion. However, a red-giant-like expansion is observed after the rapid ignition towards the end. 
\\
%\textbf{Novelty}: Our study explains the reason behind the delay in onset of helium ignition due to the difference in carbon abundance in a CO-WD.
%This is the first-ever study of the dependence of helium accreting WD evolution on its composition. We find that white dwarfs of the lower abundance of carbon accrete slightly longer before the onset of helium ignition.
\end{abstract}
%%insert keywords separated by 3 hyphens using \keywords{words}
}]
%%close the twocolumn escape here

%%include \doinum{number}for the DOI number in the header
%%include \volnum{number} for the volume number in the header
%%include \year{yyyy} for  year of publication in the header
%%include \pgrange{num--num} page range of article in the header
%%include \artcitid{num} for the article citation id
%%include \lp to print last page of the article
%%include \setcounter{page}{pagenum} for the exact starting page of the article

\doinum{}
\artcitid{\#\#\#\#}
\volnum{000}
\year{0000}
\pgrange{1--}
\setcounter{page}{1}
\lp{1}

%%%%%%%%%%%%%%%
\section{Introduction}
In many astrophysical systems, helium accretion onto a Carbon-Oxygen (CO) white dwarf (WD) plays a critical role in the binary system (Peng $\&$ Ott 2010). Most interestingly, it could be related to the progenitors of type Ia Supernovae (SNe Ia) (Ablimit 2021; Kuuttila 2021; Hillebrandt $\&$ Niemeyer 2000; Hillebrandt {\em et al.} 2000). The following are the key evolutionary pathways that led to SN Ia: 1) Semi-detached near binary stars in which WD accretes matter directly from the companion to reach the Chandrasekhar limit ($M_{Ch}$) (e.g. (Nomoto {\em et al.} 2007; Kasen $\&$ Woosley 2007; Nomoto {\em et al.} 2013; Neunteufel {\em et al.} 2016; Wang {\em et al.} 2017)]); and, 2) Explosions of sub-Chandrasekhar mass WD via "double-detonation" process. Shock waves from the He-layer at the surface of an accreting WD cause detonation (e.g. Nomoto 1982; Ropke {\em et al.} 2006).
In the case of helium accretion onto a WD, Triple-alpha (\textit{3$\alpha$}) reaction is the key to the nuclear fusion of helium into carbon. Generally, triple-alpha (\textit{3$\alpha$}) reaction occurs at ($10^8$ K) by the resonant method. However, below $10^8$ K, it can occur via a non-resonant reaction which often takes place at slow accretion rates such as \textit{$\dot{M} \lesssim 10^{-9} M_{\odot}$ $yr^{-1}$}. The study of helium accretion is crucial in understanding the helium ignition density, which determines the triggering mechanism of a supernova explosion, i.e., the explosion will occur via off-centre helium ignition or central carbon ignition (Nomoto 1982b).

Composition of WD as well as the accreted material are among critical factors that decide the characteristics of nova bursts (Starrfield {\em et al.} 2012; Schwab $\&$ Rocha 2019). 
However, the effect of the WD composition during accretion has received little attention thus far. Kumar {\em et al.} 2021 recently used MESA to investigate the effect of composition variation in a CO-WD on the slow accretion of helium-rich matter. At the bottom of the layer of accreted matter, a delay in the onset of helium ignition has been found, which is also dependent on the surface temperature of WD. Difference in ionic pressure could be one possible explanation for the delay, however, further investigation of parameters such as specific heat ($C_P$) and degeneracy parameter (\textit{$\eta$}) are required. Specific heat has implications on the cooling of the WD and on the condensation of WD core. Similarly, the electron degeneracy can play important role in explaining the temperature dependence of helium ignition (Hillman {\em et al.} 2016). 

%Effect of the WD composition has not been explored much, so far. A recent study by Kumar {\em et al.} 2021, investigated the effect of variation in carbon in a CO-WD during the slow accretion of helium rich matter using MESA. The delay in onset of helium ignition at the bottom of layer of accreted matter has been observed which also depends on the surface temperatures of WD. It seems reasonable to link the onset of helium ignition with the degeneracy pressure and the specific heat of the WD. How does the above parameters affect the ignition is the matter of study for the current paper. 

In this paper, we study the effect of carbon abundance of helium accreting WDs with a slow accretion rate on the properties like variation of specific heat which also depends on mean molecular weight (\textit{$\mu$}) and the ratio of gas pressure to the total pressure (\textit{$\beta$}); and the electron degeneracy parameter inside the accreting white dwarfs at different time scales. This paper is organised as follows: in section 2 we discuss the numerical methods used for our study, results and discussion have been presented in section 3 while section 4 contains the conclusions our study.

\section{Numerical Method}
\label{sec:method}
We have used “Modules for Experiments in Stellar Astrophysics” (MESA), which is a one-dimensional stellar evolution code (version 12778; Paxton {\em et al.} 2011; Paxton {\em et al.} 2015; Paxton {\em et al.} 2019) to study the evolution of WD during accretion. One dimensional means that the WD is spherically symmetric, and changes in WD structure occur only in the radial direction. To first order, this is an excellent approximation. For our study, we have considered CO WD of mass 0.85 $M_{\odot}$, which lies in the middle order of the mass range of WDs. Two different compositions of WDs with low and high carbon fractions have been considered, i.e., with C and O abundances (0.3, 0.7) and (0.7, 0.3), respectively. The effective temperature of WDs has also been taken in the moderate range. The parameters for the initial models of the C+O white dwarf are given in Table 1. Thus depending on the composition and effective temperature, there are four models of WDs for the study. A slow accretion rate has been chosen for study so that the triple-alpha reaction is non-resonant in all cases.
Various test suits, subroutines, functions, and control options are available in MESA to evolve stars and calculate various parameters. 
For our purpose, we have used the \texttt{make{\_}co{\_}wd} test suite to produce a WD of a given mass from a zero-age main sequence (ZAMS) 
star. We then let the WDs cool to certain specific temperatures and finally change the carbon and oxygen abundances artificially using the relax{\_}composition option in the Inlist file. For helium accretion onto the WD, the test suite wd3 has been used for our calculations. The opaqueness of the star material for the radiation is expressed in opacity, which has been taken from OPAL opacity tables. Various nuclear reactions take place during the evolution of a star as well as during the accretion. The subroutine \texttt{co{\_}burn.net} which includes 57 nuclear reactions, has been used for burning isotopes of helium, carbon and oxygen elements. Since the density of matter in the WD is very high, the pycnonuclear reactions and screening effects become important, which have been incorporated using \texttt{set{\_}rate{\_}3a} = 'FL87' (Fushiki $\&$ Lamb 1987). To get high resolution during the accretion, we set the \texttt{mesh{\_}delta{\_}coeff} which provides spatial resolution and varcontrol{\_}target that gives time resolution such that around 4000 zones of accreting white dwarfs are obtained.  We have chosen atm{\_}option = T{\_}tau atmosphere boundary condition in which temperature variation follows grey eddington relation using atm{\_}option = T{\_}tau. Using MESA we have extracted value of specific heat and degeneracy parameter in the different region of the accreting white dwarfs.

\section{Results And Discussion}
\label{sec:result}
\subsection{ Initial and Ignition Parameters}
As discussed in section-2, WDs of mass 0.85 $M_{\astrosun}$, with different abundances of C\&O, have been used for our analysis. The WDs have been generated by evolving a star of initial mass 4 $M_{\odot}$ from ZAMS followed by the main sequence and red giant phase. Then the abundances of C and O of the white dwarfs are controlled artificially by MESA, permitting them to cool to get the desired value of effective temperatures. The evolution in the WD phase is halted for two different values of effective temperature (\textit{$T_{eff}$ = 38000 K and 75000 K}). Two different compositions of these WDs have been considered, one with 30\% C \& 70\% O, and the other with 70\% C \& 30\% O. Thus, there are four models of WDs shown in Table 1, along with the numerical values of their physical parameters. 
\begin{table}[htb]
%% use tabular font for a smaller size font
\tabularfont
\caption{Parameters of white dwarf models of different abundances before accretion}
\label{table:1} %%10/12
\begin{tabular}{>{\centering\arraybackslash}m{3.5cm}>{\centering\arraybackslash}m{0.5cm}>{\centering\arraybackslash}m{0.5cm}>{\centering\arraybackslash}m{0.5cm}>{\centering\arraybackslash}m{0.5cm}}
\topline
         \hline

Parameters & A & B & C& D\\\midline
C Fraction & 0.3 & 0.7 & 0.3 & 0.7\\
\hline
Effective temperature \textit{$T_{eff} \; (10^{4} \; K)$} & 3.8 & 3.8 & 7.5 & 7.5\\
\hline
Central temperature \textit{$ T_{c}\:(10^{7}\:K)$} & 5.8 & 5.78 & 7.43 & 7.38 \\
\hline
Central density \textit{$\rho_{c}(10^{6}g$ $ cm^{-3})$} & 1.37 & 1.34 & 1.32 & 1.28 \\
\hline
 Radius (\textit{R}) $(10^{-3}R_{\odot})$ & 9.7 & 9.8 & 10.3 & 10.4 \\
\hline
\end{tabular}
\end{table}
                                              
The parameters related to helium ignition are presented in table 2, where \textit{$C_P$} and \textit{$\beta$} are the specific heat  
and ratio of gas pressure to the total pressure of the helium zone of the WD at which helium ignites. 
\textit{$\mu$} and \textit{$\eta$} are the mean molecular weight and electron degeneracy parameter in the shell at which helium ignition takes place. 
Ignition is defined as the point at which the nuclear time scale, \textit{$\tauup_{He}=\frac{C_{P}T} {\emf_{He}}$}, from helium burning 
is almost equal to $10^6$ years (Nomoto 1982). Here \textit{$C_P$} represents the specific heat, and \textit{$\emf_{He}$} is the nuclear energy generation 
rate from helium burning.     
The initial parameters related to the WD used in our study are presented in table-1. As discussed earlier, we have 
taken $T_1$ and $T_2$ as the effective temperature of WDs. Two different compositions of the CO WD have been considered, 
i.e., with carbon fraction $0.3$ and $0.7$. These are shown by models A \& B for relatively cool and C \& D for relatively 
higher temperatures. Core temperature and densities have been shown in the table-1. 
The WDs have been subjected to the accretion of helium-rich matter with two different accretion rates. 
One of the accretion rate is as slow as $5 \times 10^{-10}$ \textit{$M_{\odot}$} $yr^{-1}$. In this case, the triple-alpha 
reaction takes place via the non-resonant method. The other accretion rate is high ($5 \times 10^{-8}$ \textit{$M_{\odot}$} $yr^{-1}$) 
in which triple-alpha reaction occurs via resonance. 
%\vspace{-2em}

%%%%%%%%%%%
\begin{table}[htb]
%% use tabular font for a smaller size font
\tabularfont
\caption{He Ignition Parameters of white dwarfs at 38000 K  with accretion rate $5 \times 10^{-10}$ $M_{\odot}$ $yr^{-1}$}
\label{table:2} %%10/12
\begin{tabular}{>{\centering\arraybackslash}m{3.5cm}>{\centering\arraybackslash}m{2.0cm}>{\centering\arraybackslash}m{2.0cm}}
\topline
\hline
Parameters & A & B \\\midline

\textit{t} (10$^8$ yr) & 9.627 & 9.565\\
\hline
\textit{$C_P \:(10^7 \; ergs \; g^{-1} \; K^{-1})$} & 4.155 & 4.130  \\
\hline
\textit{$(1-\betaup) \: (10^{-9})$} & 0.43 & 0.56 \\
\hline
\textit{$\mu$} & 1.33348653 & 1.33342511 \\
\hline
\textit{$Degeneracy \; Parameter \; (\eta) $} & 388.20 & 360.60 \\
\hline
\end{tabular}
%%use \tablenotes{footnote} to get the table foot note
%\tablenotes{Sample table footnote}%%9/11
\end{table}
\begin{table}[htb]
%% use tabular font for a smaller size font
\tabularfont
\caption{He Ignition Parameters of white dwarfs at 75000 K $5 \times 10^{-10}$ $M_{\odot}$ $yr^{-1}$}
\label{table:3} %%10/12
\begin{tabular}{>{\centering\arraybackslash}m{3.5cm}>{\centering\arraybackslash}m{2.0cm}>{\centering\arraybackslash}m{2.0cm}>{\centering\arraybackslash}m{0.5cm}>{\centering\arraybackslash}m{0.5cm}}
\topline
         \hline
Parameters & C& D\\\midline

\textit{t} (10$^8$ yr) & 9.645 & 9.573 \\
\hline
\textit{$C_P \:(10^7 \; ergs \; g^{-1} \; K^{-1})$} & 1.700 & 1.783 \\
\hline
\textit{$ (1-\betaup) \: (10^{-8})$} & 0.1 & 0.16 \\
\hline
\textit{$\mu$} & 1.73631547  &1.71684016 \\
\hline
\textit{$Degeneracy \; Parameter \; (\eta) $} & 317.22& 284.04 \\
\hline
\end{tabular}
%%use \tablenotes{footnote} to get the table foot note
%\tablenotes{Sample table footnote}%%9/11
\end{table}

%%%%%%%%%%%%%%%%%%
\subsection{Profile of Specific heat \textit{C$_P$}}
\begin{figure}[!t]
  \centering\includegraphics[width=1.0\columnwidth]{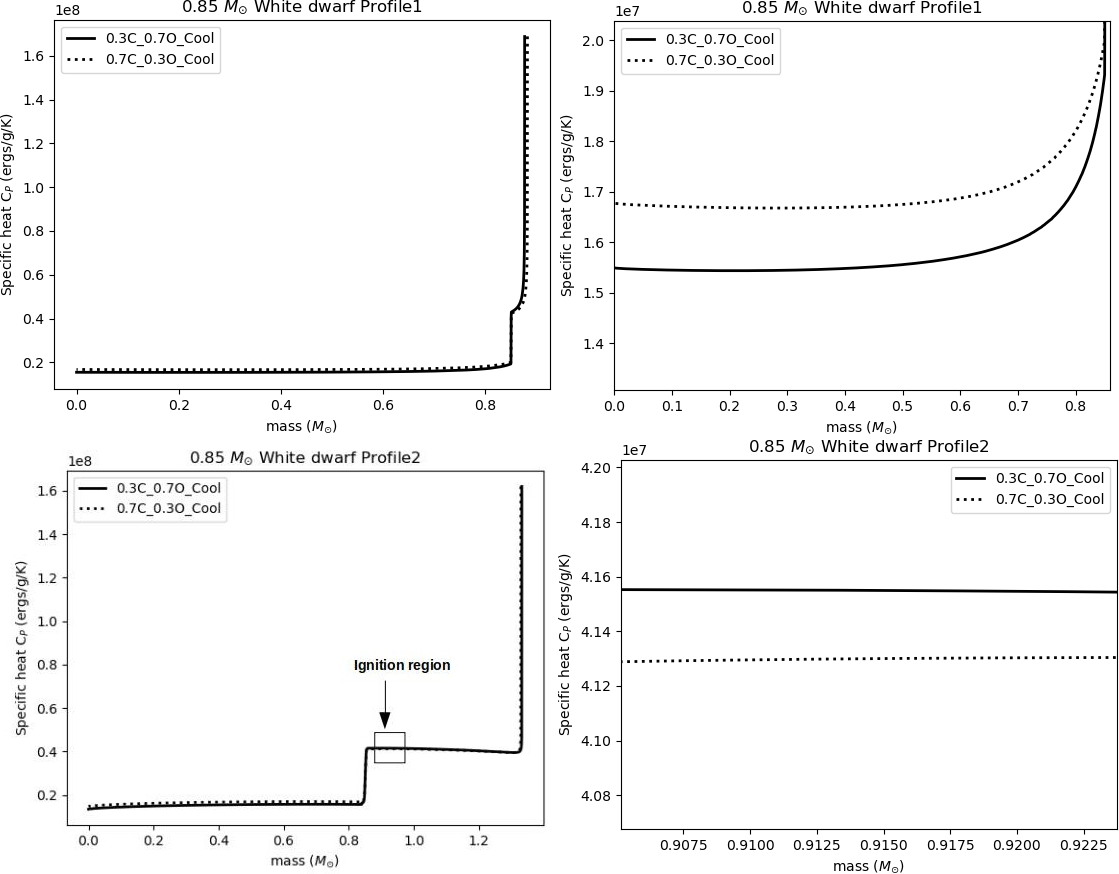}
\caption{Profile of specific heat $C_P$ inside the accreting white dwarfs having initial temperature 38000K. Top left: Profile of $C_P$ in the beginning. Top right: zoom portion up to the core of the WD. Bottom left: Profile of $C_P$
at the time when helium ignition takes place. Bottom right: Variation of $C_P$ in the zone of helium ignition.}
\label{fig:1}
\end{figure}

\begin{figure}[!t]
  \centering\includegraphics[width=1.0\columnwidth]{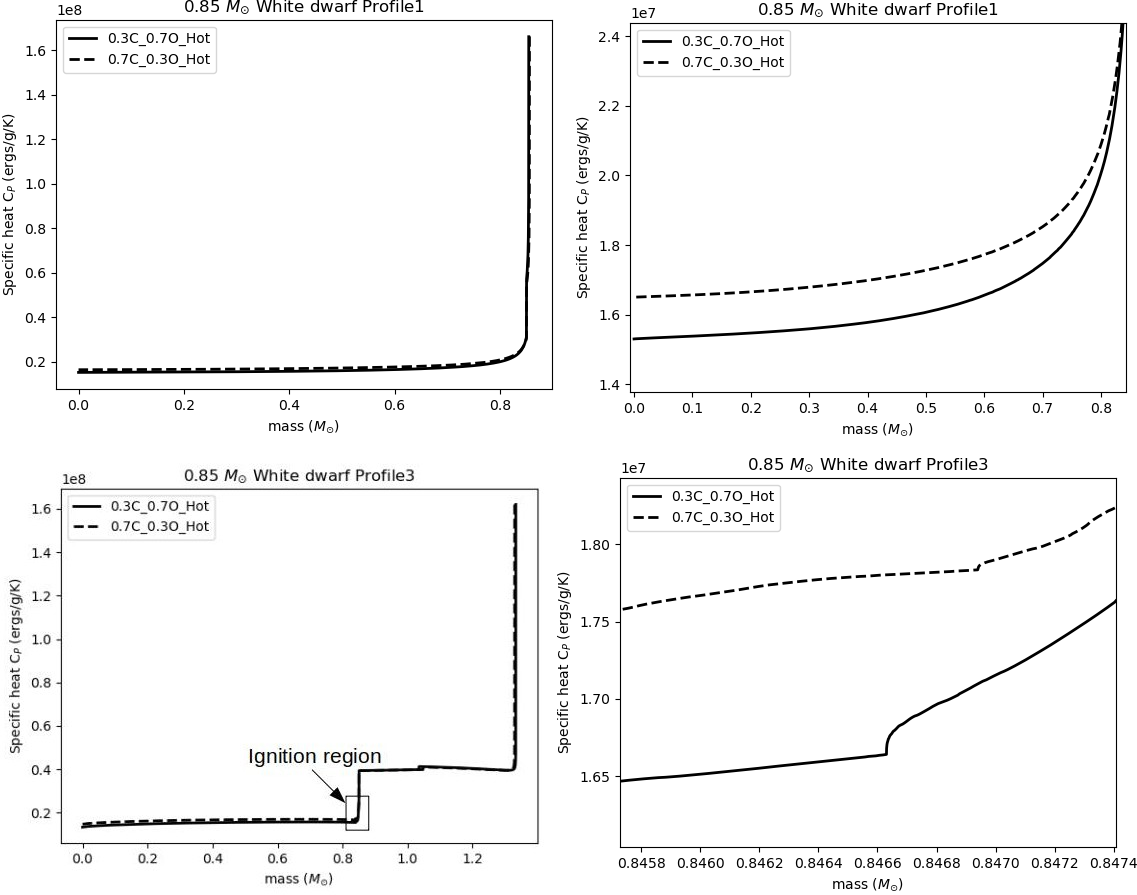}
\caption{Same as in Fig 1 profile of specific heat C$_P$ inside the accreting white dwarfs of different abundances having initial temperature 75000K }
\label{fig:2}
\end{figure}

Since specific heat is a derivative of internal energy, \textit{U}, calculating specific heat necessitates an understanding of internal energy. In the case of WD contribution due to radiation field can be ignored for internal energy. Thus the main components of internal energy are the degenerate electron gas and the perfect ion gas. The specific heat inside the accreting white dwarf has been calculated using the following equation ( see chapter 16, Weiss {\em et al.} 2012) 
\begin{equation}\label{eq1}
       \mathit{C_P = \frac{5}{2} \left [ \frac{P \delta}{\rho T}\right ]}
\end{equation}
Where \textit{$P$} is the total pressure that also includes electron degeneracy pressure, \textit{$\delta\; \approx \; \frac{3}{2}\frac{\mu_e}{\mu_o} \frac{1}{\eta} $} is a parameter that depends on mean molecular weight per electron (\textit{$\mu_e$}), mean molecular weight (\textit{$\mu$}), and electron degeneracy parameter (\textit{$\eta$}).
Figure~\ref{fig:1} and \ref{fig:2} show the variation of C$_P$ in different zones of the WD.
It is clear from the top panel of both the figures that in the inner region the specific heat is small and is almost constant. Also, \textit{$C_P$} is smaller inside the WD with lower carbon fraction in both low and high temperature WDs. We also notice that \textit{$C_P$} increases rapidly as we move from center to the surface of white dwarfs. 
Time variation of \textit{C$_P$} is quite interesting in the WD with lower temperature. At the initial stage, specific heat is relatively larger in 0.7C white dwarf as compared to white dwarf of 0.3 C. However, the situation is reversed by the time of ignition as can be seen from the bottom panel of figure 1. This explains the slight delay in helium ignition in the WD with lower carbon fraction (Kumar {\em et al.} 2021)\cite{Kumar2021}.
%As the accretion of helium starts, specific heat is more in white dwarf of o.7 C as compared to white dwarf of 0.3 C. but at the time of ignition specific heat increases more rapidly in white dwarf of 0.3 C that indicates that helium ignition in the bottom of accreted helium on to a WD with 0.3 C starts later as compared to the WD with 0.7 C.

%%%%%%%%%%%%%%%%%%%%%%%%%%%%%%%%%%%%%%%%%%%%%%
 
%%%%%%%%%%%%%%%%%%

\subsection{Profile of Degeneracy Parameter {\textit{$\eta$}}}
\begin{figure}[!t]
  \centering\includegraphics[width=8cm,angle=0]{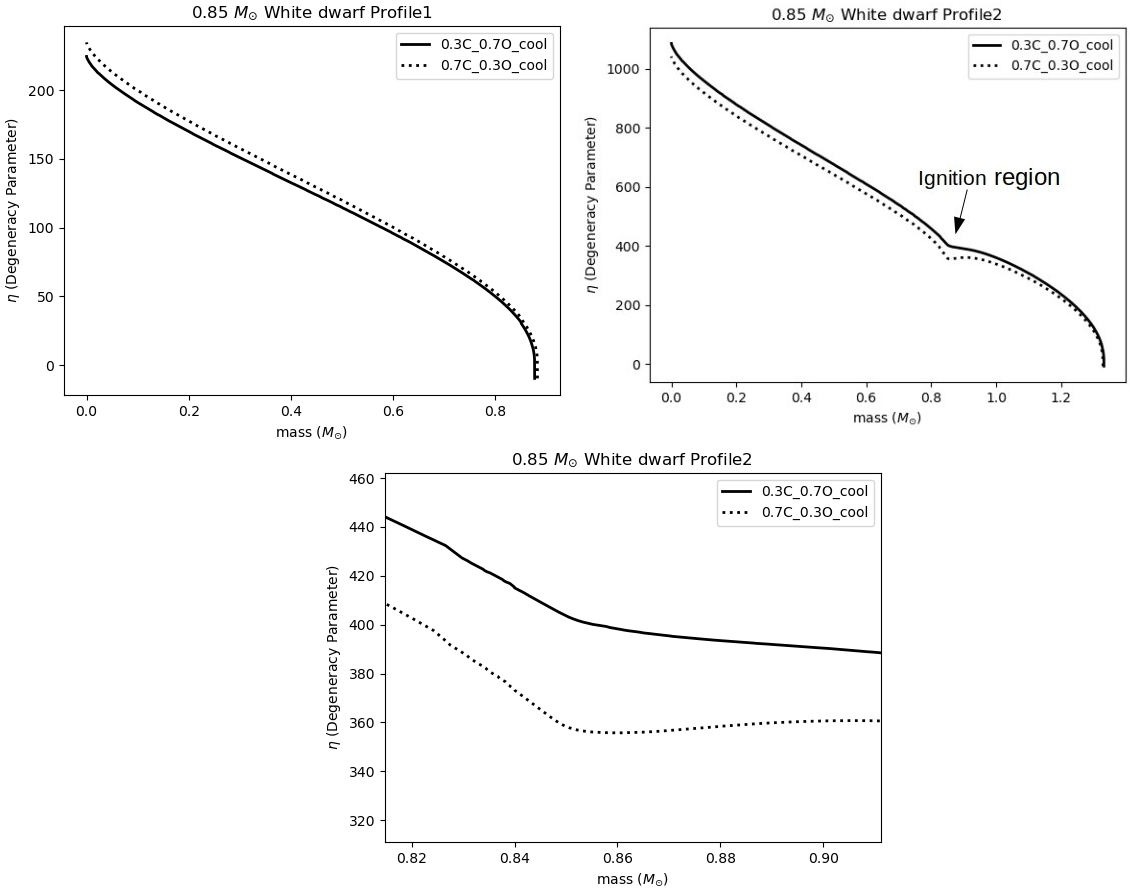}
\caption{The upper left and upper right panels provide the profiles of electron degeneracy parameter inside the white dwarfs of different oxygen and carbon abundances but at the same initial effective temperature 38000K after some times of accretion and at the time of helium ignition respectively. The bottom panels is the zoom portion of the zone of helium ignition. }
\label{fig:3}
\end{figure}

\begin{figure}[!t]
  \centering\includegraphics[width=8cm,angle=0]{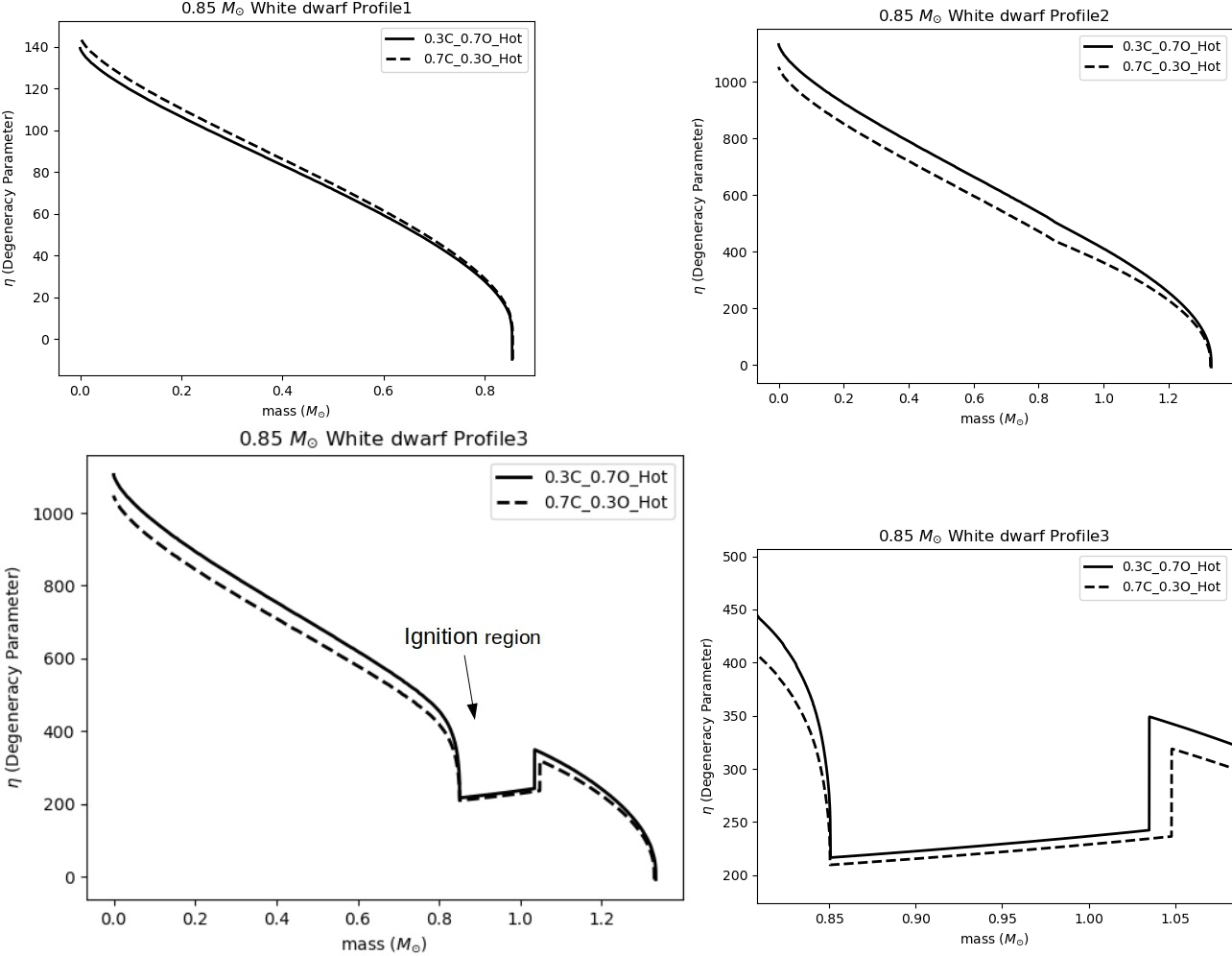}
\caption{The top left, right, and bottom left panels show the  profiles of electron degeneracy parameter inside the white dwarfs of different oxygen and carbon abundances at initial effective temperature 75000K at the time of beginning of accretion, after one billion of accretion 
and at the time when helium ignites. And the bottom right panel is the zoom portion of ignited region.}
\label{fig:4}
\end{figure}

The electron degeneracy parameter inside the accreting white dwarfs from center to surface is calculated by (see chapter 15, Weiss {\em et al.} 2012) 
\begin{equation}\label{eq2}
 \mathit{exp \; (\eta) = \frac{h^{3} n_e}{2(2\pi m_{e} k T)^{3/2}}}
 \end{equation}
Here \textit{$\eta$} is electron degeneracy parameter that depends on electron density $n_e$ and temperature \textit{T}. The variation of \textit{$\eta$} is shown in figure~\ref{fig:3} and \ref{fig:4}. The initial stage of the cool WD is shown as profile 1 in figure~\ref{fig:3} which  indicates that the degeneracy is high at the center and it drops to zero as one moves towards the surface of WD. It should be noted that in profile 1 the level of degeneracy in 0.7C WD is slightly higher than that of 0.3C. The variation of degeneracy parameter at the time of helium ignition is shown as profile 2 in figure~\ref{fig:3}. Contrary to the initial stage the degeneracy parameter is higher in 0.3C instead of 0.7C. The dip in the graph shows the helium ignition region. Almost similar trend is seen in figure 4 which represents the case of hot WD. 

%As the degeneracy pressure is very high at the center of the white dwarfs and it decreases towards the surface. From Fig 3., Profile1 shows that most of the region of the white dwarfs is highly degenerate and it is seen that at the time of accretion of helium starts degeneracy pressure of white dwarf having 0.3 C is slightly less than the degeneracy pressure of WD with 0.7 C. but at the time of ignition of helium in the bottom of accreted helium on to the white dwarfs of different abundances, white dwarf with high carbon abundance has low degeneracy pressure as compared to the WD with small carbon abundances.

%%%%%%%%%%%%%%%%%%%%%%%%%%%%%%%%%%%%%%%%%%%%%

%%%%%%%%%%%%%%%
%%%%%%%%%%%%%%%

%%%%%%%%%%%%%%%%%

%%%%%%%%%%%%%%%
%\begin{figure}[!t]
 % \includegraphics[width=.8\columnwidth]{Images/FIg 6_Kumar.png} 
%\caption{Variation in effective temperature $T_{eff}$ during the accretion phase of helium accreting white dwarf having 75000 K initial effective temperature.}
%\label{fig:6}
%\end{figure}

%%%%%%%%%%%%%%%

%%%%%%%%%%%%%%%%%

%%%%%%%%%%%%%%

%\vspace{-2em}
%%Use table environment for a table in one column

%\begin{table}[htb]
%% use tabular font for a smaller size font
%\tabularfont
%%\caption{Initial And Ignition Parameters}\label{tableExample} %%10/12
%\begin{tabular}{lcccccc}
%\topline
%Parameters & A & B & C& D& E& F\\\midline
%C and O &0.3&0.5&0.7&0.3&0.5&0.7\\
%$T_{eff}(10^{4}K)$&3.7&3.7&3.7&5.2&5.2&5.2\\
%$T_{c}(10^{7}K)$&5.4&5.4&5.5&6.1 &6.1 &6.1 \\
%$\rho_{c}(10^{6}g cm^{-3})$&3.5&3.5&3.5&3.4&3.4&3.4 \\

%\hline
%\end{tabular}
%%use \tablenotes{footnote} to get the table foot note
%\tablenotes{Sample table footnote}%%9/11
%\end{table}
%%%%%%%%%%%%%%%%%%

%%%%%%%%%%%%%%%%%%%%%%
%%%%%%%%%%%%%%%%%%%

%%Use table* environment to get the table spanning both the columns

%%An example of a figure

%\begin{figure}[!t]
%\includegraphics[width=.8\columnwidth]{fig1.eps}
%\caption{caption goes here}\label{figOne}
%\end{figure}

%%An example of a double column figure
%%Use figure* environment

%\begin{figure*}
%\centering\includegraphics[height=.15\textheight]{fig1.eps}
%\caption{caption spanning two columns}
%\centering\includegraphics[height=.25\textheight]{fig1.eps}
%\caption{caption here}
%\end{figure*}

%\vspace{-2em}
\section{Conclusion}
\label{sec:conclusion}
In the present article we have discussed the variation of specific heat, $C_P$, and degeneracy parameter, $\eta$, inside the accreting WDs. Different effective temperatures and different carbon abundances have been considered. It is found that specific heat is discontinuous near the bottom of accreted helium layer which is expected at the onset of helium ignition. The degeneracy parameter is comparatively smaller for hot WD which is consistent with ((Hillman {\em et al.} 2016)). Also, the degeneracy parameter has a dip near the ignition region. 
We notice that at the initial stage the specific heat is lower for the WD with higher carbon fraction. This is independent of the effective temperature and is valid throughout the whole of WD. However, the situation is different at the later stages of evolution. When the helium ignition takes place in the relatively cool WD the specific heat near the accreted layer becomes higher for higher carbon fraction. On the other hand, no such effect has been noticed in the hot WD. This behaviour might explain the delay in onset of helium ignition in cool WD with lower carbon abundance (Kumar {\em et al.} 2021). 

%In the present article, we have discussed the variation of specific heat \textit{C$_P$} from center to surface inside the accreting white dwarfs with different carbon oxygen: (0.3 C, 0.7 O) and (0.7 C, 0.3 O) having two different temperatures 38000 K (cool) and 75000 K (hot). In case of cool white dwarfs, It is found that in the deep interior of the white dwarf with 70$\%$ carbon abundance has high specific heat \textit{$C_P$} as compared to the white dwarf of 30$\%$ carbon composition with oxygen while in the region of white dwarf where helium ignition taking place, \textit{$C_P$} is high in WD with 30$\%$ carbon composition as compared to the WD with 70 $\%$. In case of hot white dwarf, specific heat \textit{$C_P$} is high in the interior as well as in region of helium ignition of the white dwarf of 70$\%$ carbon composition as compared to the WD with 30$\%$ carbon abundances. There is a dip in the electron degeneracy parameter \textit{$\eta$} at the helium ignition region.

%%%%%%%%%%%%%%%%%%%%%%
%\section{An appendix section}
%Text goes here (Radhakrishnan 1980).
%\begin{equation}
%x=a+b+c
%\end{equation}

%\vspace{-2em}

%\section{Another appendix section}
%Text goes here.
%\begin{equation}
%y^2=ax+b+c
%\end{equation}
%\vspace{-3em}

%%Use section* for acknowledgements
\section*{Acknowledgements}
\vspace{-0.5em}
SG thanks SERB for financial assistance (EMR/2017/003714).

%%use \balance somewhere in the left column of the last page to balance the two columns in the end page
%\bibliography{references}

\end{document}